\documentclass[a4paper,fleqn,usenatbib]{mnras}

\usepackage[T1]{fontenc}
\usepackage{ae,aecompl}

\usepackage{graphicx}   
\usepackage{amsmath}    
\usepackage{amssymb}    

\title[New Insight into Particle Reservoirs]{New Insight into the Formation Mechanism of the Energetic Particle Reservoirs in the Heliosphere}

\author[H.-Q. He]{H.-Q. He$^{1,2,3}$\thanks{E-mail: hqhe@mail.iggcas.ac.cn (H.-Q. He)} \\
$^{1}$ Key Laboratory of Earth and Planetary Physics, Institute of
Geology and Geophysics, Chinese Academy of Sciences, Beijing 100029,
China \\
$^{2}$ Innovation Academy for Earth Science, Chinese Academy of
Sciences, Beijing 100029, China \\
$^{3}$ Beijing National Observatory of Space Environment, Institute
of Geology and Geophysics, Chinese Academy of Sciences, Beijing
100029, China}

\date{Accepted 2021 August 4. Received 2021 August 4; in original form 2021 May 2}

\pubyear{2021}

\begin{document}
\label{firstpage}
\pagerange{\pageref{firstpage}--\pageref{lastpage}}
\maketitle

\begin{abstract}
The concept of energetic particle reservoirs, essentially based on
the assumption of the presence of outer reflecting
boundaries/magnetic mirrors or diffusion barriers (deterministic)
rather than on the effect of particle diffusive propagation
(stochastic) in magnetic turbulence, has been used for decades to
describe the space-extended decay phases of energetic particle
events within the fields of space physics, solar physics, and plasma
physics. Using five-dimensional time-dependent Fokker-Planck
transport equation simulations, in this work we demonstrate that the
so-called particle reservoirs are naturally explained and
quantitatively reproduced by diffusion processes in turbulent
magnetic fields, without invoking the hypothesis of reflecting
boundaries. Our results strongly suggest that the so-called
``reservoir'' (based on deterministic structure) should be renamed
``flood'' (based on stochastic diffusion), which symbolizes an
authentic shift in thinking and in pragmatic rationale for the
studies of energetic particles and relevant plasma phenomena in
heliophysics and in astrophysics.
\end{abstract}




\section{Introduction}
Solar energetic particles (SEPs), emitted by powerful solar bursts,
can serve as sample particles for cosmic ray studies by providing
fundamental information regarding particle acceleration and
propagation in the turbulent magnetic fields. Moreover, SEPs risk
astronauts and damage satellites in space, and significantly
influence the solar-terrestrial space environment, space weather,
and space climate. Hence, the subject of energetic particles has
become one of the most important foci in space physics, solar
physics, plasma physics, and astrophysics. Basically, the
propagation of energetic particles in the turbulent magnetic fields
consists of a handful of fundamental mechanisms, such as
field-aligned particle streaming, convection with the plasma medium,
adiabatic focusing, adiabatic deceleration, pitch-angle diffusion,
and perpendicular diffusion. The concept of particle diffusion
process at microscale is essentially based on the philosophy and
thought of stochasticism (probabilism) in a mathematically tractable
style \citep[e.g.][]{Brown1828,Kubo1957,Parker1965,Jokipii1966}. The
energetic charged particles in the turbulent magnetic fields undergo
significant diffusion processes due to the effects of stochastic
magnetic fluctuations. The parallel diffusion along the guide
magnetic field has been intensely and extensively studied; however,
the perpendicular diffusion across the guide magnetic field has
remained a puzzle for quite a long time. Recently, the important
effect of the perpendicular diffusion mechanism in the transport
processes has been found
\citep[e.g.][]{Zhang2003,Matthaeus2003,Shalchi2004,He2015}. The
effects of nonlinear terms have been introduced and discussed so as
to explain perpendicular diffusion and scattering of cosmic rays
near $90^{\circ}$ pitch-angle.

There is a well-known phenomenon or conception in the field of
energetic particles, especially in the SEP community, i.e., the
``reservoir''
\citep[e.g.][]{Simnett1995,Poletto2004,Klecker2006,Reames2013,Desai2016,Reames2017,Simnett2017,Reames2018,Anastasiadis2019}.
The reservoir effect refers to the interesting phenomenon where the
flux-time profiles of particles evolve similarly in time with almost
the same declining rate in the decay stage of the particle events.
This intriguing feature was first reported from the
wide-heliolongitude ($\sim180^{\circ}$) concurrent observations of
$\sim20$ MeV protons measured by the spacecraft Interplanetary
Monitoring Platform (IMP) 4 and Pioneer 6 and 7
\citep{McKibben1972}. Twenty years later, uniform intensities were
observed spanning $\gtrsim2.5$ AU in the radial direction between
spacecraft IMP 8 and Ulysses, and thus were named ``reservoirs'' to
illustrate the uniformity of particle distribution
\citep{Roelof1992}. The particle (protons, electrons, and
heavy-ions) reservoirs are often seen by widely separated spacecraft
at very different heliolongitudes, heliolatitudes, and radial
distances
\citep[e.g.][]{Reames1996,Maclennan2001,Dalla2003,Sanderson2003,Daibog2003,McKibben2003,Lario2003,Lario2010}.
Basically, the notion of reservoir is based solely on the
hypothetical presence of deterministic magnetic structures at
macroscale, such as diffusion barriers, outer reflecting boundaries
or magnetic mirrors produced by plasma disturbances in the solar
wind. This idea invokes the presence of such diffusion barriers or
reflecting boundaries to contain the particles long enough to
gradually and uniformly redistribute them in heliographic azimuth
and latitude. In this picture, the so-called reflecting boundaries
or diffusion barriers are assumed to exist in the space to play the
role of ``reservoir dam'' to impede, reflect, and redistribute
energetic particles within the reservoirs. Ultimately, the concept
of ``reservoir'' originates from conventional thought of
scatter-free (deterministic) propagation, meanwhile ignoring the
important effects of diffusive (stochastic) transport
\citep[e.g.][]{Roelof1975,Nolte1975}.

Recently, multi-spacecraft SEP observations, achieved with the
spacecraft STEREO-A/B, ACE, SOHO and Wind, have shown that SEPs
originating from a limited solar source can transport to very
distant heliographic longitudes as wide as $136^{\circ}-360^{\circ}$
both in the so-called gradual and impulsive ($^{3}$He-rich) events
\citep[e.g.][]{Dresing2012,Wiedenbeck2013,Richardson2014,Cohen2014,Gomez-Herrero2015}.
Essentially, these wide longitudinal spreads of SEP events are
identical to the SEP ``reservoir'' phenomenon, since both of these
two SEP effects indicate the same manifestation: the broad and
gradually uniform particle distribution in the heliosphere.
Therefore, the key physical mechanism for fundamentally
understanding these novel observations still remains a debatable
conventional issue regarding scatter-free propagation (determinism)
and diffusive transport (randomness). To quantitatively explain and
reproduce these intriguing particle phenomena in a compelling
fashion eagerly requires a thorough and detailed investigation by
means of physics-based numerical simulations and computationally
tractable descriptions of particle transport in three-dimensional
turbulent interplanetary magnetic fields (IMF).

\section{Fokker-Planck Model}
We present the physics-based multi-dimensional Fokker-Planck
transport model and relevant numerical calculations to shed new
light on the formation mechanism of reservoirs of energetic
particles in the heliosphere. The Fokker-Planck model is based upon
dynamical stochastic diffusion motions of charged energetic
particles in magnetic fields with turbulent effects, and can be
formatted as
\citep[e.g.][]{Schlickeiser2002,Parker1965,Zhang2009,He2011,Droge2016}
\begin{eqnarray}
{}&&\frac{\partial f}{\partial t}+\mu v\frac{\partial f}{\partial
z}+{\bf V}^{sw}\cdot\nabla f+\frac{dp}{dt}\frac{\partial f}{\partial
p}+\frac{d\mu}{dt}\frac{\partial f}{\partial \mu}  \nonumber\\
{}&&-\frac{\partial}{\partial\mu}\left(D_{\mu\mu}\frac{\partial
f}{\partial \mu}\right)-\frac{\partial}{\partial
x}\bigg(\kappa_{xx}\frac{\partial f}{\partial x}\bigg) \nonumber\\
{}&&-\frac{\partial}{\partial y}\left(\kappa_{yy}\frac{\partial
f}{\partial y}\right)=Q({\bf x},p,t),  \label{transport-equation}
\end{eqnarray}
where $f(\textbf{x},\mu,p,t)$ denotes the gyrophase-averaged
distribution function of particles, $\textbf{x}$ is spatial position
of particles, $z$ is coordinate along magnetic field line, $p$ is
particle momentum, $\mu$ is particle pitch-angle cosine, $t$ is
time, $v$ is particle velocity, $\textbf{V}^{sw}$ is solar wind
velocity, $\kappa_{xx}$ and $\kappa_{yy}$ are perpendicular
diffusion coefficients of particles, $dp/dt$ represents adiabatic
deceleration effect, $d\mu/dt$ represents magnetic focusing effect
and solar wind flow divergence, and $Q(\textbf{x},p,t)$ is source
term of particles. We use the technique of time-backward Markov
stochastic processes to solve the above five-dimensional
Fokker-Planck model \citep{Zhang2009,He2011}. A constant solar wind
speed of $400~km~s^{-1}$, a Parker-type magnetic field with
magnitude $B=5$ nT at $1$ AU, and a particle source with limited
longitudinal and latitudinal coverage centered at $0^{\circ}$
heliolatitude (Equator) are typically used in the simulations. We
simulate $6\times10^{7}-1.2\times10^{8}$ particles on a
supercomputer cluster. We analyze statistical data from numerical
simulations and obtain intensity-time profiles of SEP events with
different features. An arbitrary unit for particle flux is
conveniently used in plotting figures. It is important to note that
the simulation results and conclusions shown here essentially hold
for varying values of the model parameters.

\section{Results}
\begin{figure}
    \centering
    \includegraphics[width=\columnwidth]{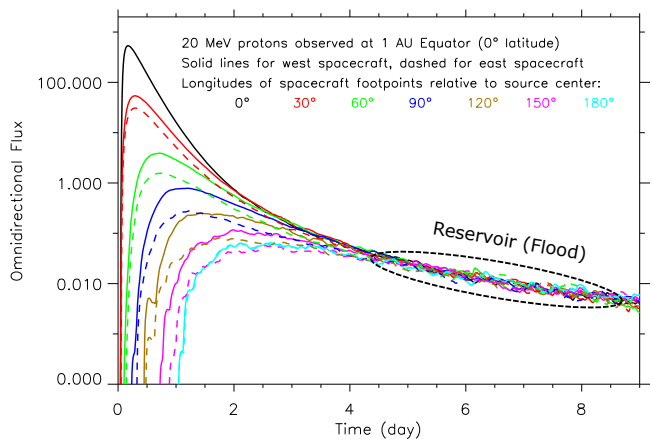}
    \caption{Numerical simulations of $360^{\circ}$-longitude distribution
and reservoir (flood) effect of $20$ MeV protons. The flux-time
profiles are obtained by numerically solving five-dimensional
Fokker-Planck focused transport equation incorporating diffusion
processes. A typical circumsolar particle reservoir spanning
$360^{\circ}$ in longitude is reproduced without invoking a
hypothetical outer reflecting boundary.}
    \label{proton-20MeV-lon}
\end{figure}

The reservoirs were first noted in the wide-heliolongitude
observations of $\sim20$ MeV protons \citep{McKibben1972}. Fig.
\ref{proton-20MeV-lon} presents our numerical simulations of
flux-time profiles of $20$ MeV protons transporting in the
interplanetary space. The diffusion coefficients of energetic
particles are set as follows: the radial mean free path
$\lambda_{r}=0.32$ AU (corresponding to the parallel mean free path
$\lambda_{\parallel}=0.64$ AU at 1 AU), and the perpendicular mean
free paths $\lambda_{x}=\lambda_{y}=0.008$ AU. We note that the
values of the mean free paths chosen for numerical simulations in
this work are consistent with the observational and theoretical
results within the SEP community
\citep[e.g.][]{Bieber1994,Droge2000,Matthaeus2003,Bieber2004,He2012a,He2012b,He2013}.
The observers (spacecraft) are located at 1 AU Equator ($0^{\circ}$
heliolatitude). The different longitudinal separations between the
particle source and the magnetic field line footpoints of different
observers are: $0^{\circ}$, $30^{\circ}$, $60^{\circ}$,
$90^{\circ}$, $120^{\circ}$, $150^{\circ}$, and $180^{\circ}$. In
Fig. \ref{proton-20MeV-lon}, the solid and dashed lines denote the
flux-time observations made by spacecraft located at west and east,
respectively, relative to magnetic field line connecting the center
of the source. As one can see, in the beginning, the farther the
magnetic footpoint of the observer is away from the particle source,
the smaller is the particle flux observed and also the later the
onset and the peak flux of the event appear \citep{He2011}. In
addition, with the same heliolongitude separation between particle
source and magnetic footpoints of observers, the particle fluxes
observed by spacecraft located at western side of source are
systematically larger than the fluxes observed by spacecraft located
at eastern side, and also the SEP event (onset and peak) arrives at
the western spacecraft earlier than at the eastern spacecraft
\citep{He2011,He2017}. Most interestingly, however, in the late
phase of SEP events, the particle fluxes observed by these widely
longitudinally separated spacecraft show nearly the equal
intensities and decay at the same rate. Therefore, the typical
features of a circumsolar particle reservoir which spans
$360^{\circ}$ in heliolongitude at $1$ AU have been successfully
reproduced in the numerical simulations based upon five-dimensional
Fokker-Planck focused transport equation incorporating parallel
diffusion and perpendicular diffusion. It is noteworthy that in the
simulations, no diffusion barriers (reflecting boundaries/magnetic
mirrors) have been assumed and employed to play the role of
``reservoir dam'' to reflect/redistribute energetic particles. That
is to say, the traditional hypothesis of reflecting boundaries needs
not to be necessarily invoked in the numerical reproduction of
energetic particle reservoirs. Actually, contrary to the
expectations of the traditional thoughts, the presence of outer
reflecting boundaries, diffusion barriers, or magnetic mirrors
(magnetic bottles) will prevent the uniform distribution and
reservoir formation of energetic particles to occur for the
observers at different locations in the interplanetary medium, which
has recently been confirmed by \citet{Wijsen2020}.

\begin{figure}
    \centering
    \includegraphics[width=\columnwidth]{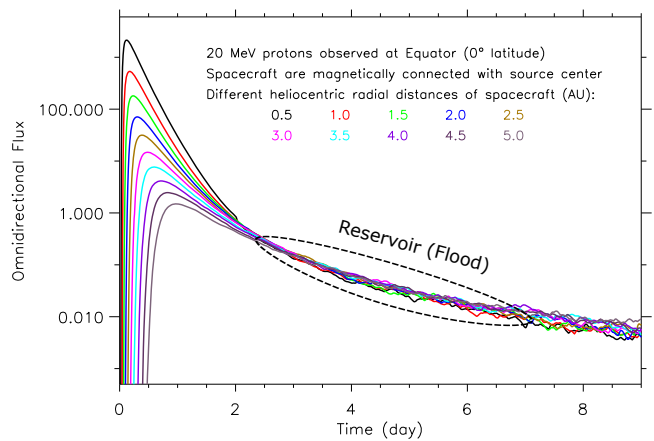}
    \caption{Numerical simulations of radial-spread distribution and
reservoir (flood) effect of $20$ MeV protons in inner and outer
heliosphere. A typical radial particle reservoir spanning inner and
outer heliosphere is reproduced without invoking the artificial
hypothesis of outer reflecting boundaries or diffusion barriers.}
    \label{proton-20MeV-radial}
\end{figure}

The particle reservoirs are observed radially spreading out over
several AU between IMP 8 near Earth and Ulysses in the outer
heliosphere \citep[e.g.][]{Roelof1992,Lario2010}. Fig.
\ref{proton-20MeV-radial} presents the numerical simulation results
of radial distribution of $20$ MeV protons in the inner and outer
heliosphere. The diffusion coefficients of particles are the same as
those used in Fig. \ref{proton-20MeV-lon}. The observers here are
located in the Sun's equatorial plane and are magnetically connected
with the center of the particle source through a nominal Parker
spiral. The different heliocentric radial distances of different
observers are: $0.5$, $1.0$, $1.5$, $2.0$, $2.5$, $3.0$, $3.5$,
$4.0$, $4.5$ and $5.0$ AU. As one can see, during the early phases
of particle events, the farther the observer is radially away from
the solar source, the smaller is the particle intensity detected and
also the later the onset and the peak intensity of the event occur.
However, during the late phases of particle events, all the
intensity profiles present nearly the same intensities and decay
rates within the range of statistical errors. Hence the radial
particle reservoir has been reproduced in the Fokker-Planck equation
simulations without invoking the hypothetical reflecting boundaries
or diffusion barriers.

\begin{figure}
    \centering
    \includegraphics[width=\columnwidth]{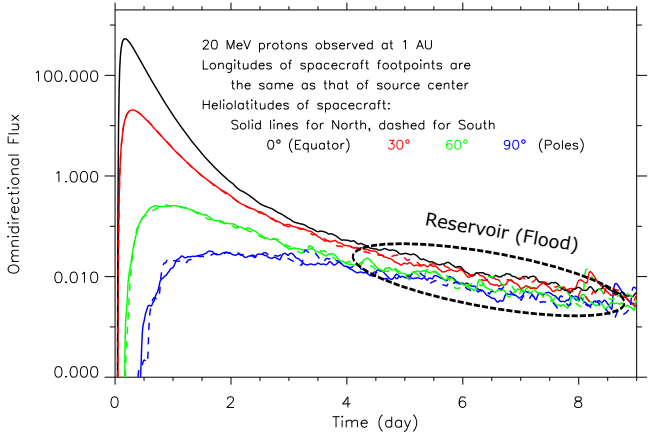}
    \caption{Numerical simulations of $180^{\circ}$-latitude distribution
and reservoir (flood) effect of $20$ MeV protons. A typical
latitudinal particle reservoir extending to the highest latitudes of
$90^{\circ}$ (magnetic poles), both North and South, is
quantitatively reproduced without invoking the hypothetical outer
reflecting boundaries.}
    \label{proton-20MeV-lat}
\end{figure}

The SEP reservoirs are also found extending to quite high
heliolatitudes, e.g., up to $>70^{\circ}$, both North and South
\citep[e.g.][]{Dalla2003,Sanderson2003,McKibben2003,Lario2003}. Fig.
\ref{proton-20MeV-lat} displays the simulation results of
latitudinal distribution of $20$ MeV protons. The parallel and
perpendicular diffusion coefficients of particles are the same as
those used in Fig. \ref{proton-20MeV-lon}. The observers are located
at 1 AU. The heliolongitudes of magnetic footpoints of observers are
the same as the heliolongitude of the center of particle source. The
different heliolatitudes of different observers are: $0^{\circ}$
(Equator), $30^{\circ}$, $60^{\circ}$, and $90^{\circ}$ (Poles). The
solid and dashed curves in Fig. \ref{proton-20MeV-lat} denote the
flux-time profiles of SEPs observed in the northern and southern
hemispheres, respectively. We can see that during the beginning
stage of SEP events, the farther the magnetic footpoint of observer
is away from the particle source in latitude, the smaller is the SEP
flux observed and also the later the SEP event arrives at the
observer's position. Due to the important effect of perpendicular
diffusion, even the observers at the highest heliolatitudes of
$90^{\circ}$ (magnetic poles), both North and South, can detect
considerable particle fluxes. Different from the East-West
longitudinal asymmetry of particle distribution shown in Fig.
\ref{proton-20MeV-lon}, there exists no North-South latitudinal
asymmetry phenomenon of particle intensity within the range of
statistical errors. The reason is that the geometry of the Parker
IMF is North-South latitudinally symmetric. During the late stage of
the particle events, all the particle fluxes observed by these very
widely latitudinally separated spacecraft show nearly equal fluxes
(to within a small factor of $\sim2-3$) that decline similarly with
time. Therefore, the simulations based on five-dimensional
Fokker-Planck focused transport equation have quantitatively
reproduced the latitudinal particle reservoirs in the heliosphere.
See Appendix A for additional simulation results and some
comparisons with multispacecraft observations, and Appendix B for
some implicating remarks. Note that both Appendices A and B are in
the Supplementary Material (Supporting Information).

The five-dimensional Fokker-Planck focused transport model
incorporating both parallel and perpendicular diffusion has provided
us an essential basis and a very useful tool to better understand
the three-dimensional propagation of energetic particles in the
heliosphere filled with turbulent IMF. By numerically solving the
Fokker-Planck model on a supercomputer cluster, we have simulated
and reproduced a number of energetic particle reservoirs under
various physical conditions and in various physical scenarios. We
point out that altering the values of the model parameters does not
qualitatively influence the reproduction of particle reservoirs. The
simulations based on five-dimensional Fokker-Planck focused
transport equation help us obtain quantitative and physical insights
into the realistic formation mechanism of the so-called energetic
particle reservoirs.

\section{Discussion and Summary}
\begin{figure}
    \centering
    \includegraphics[width=\columnwidth]{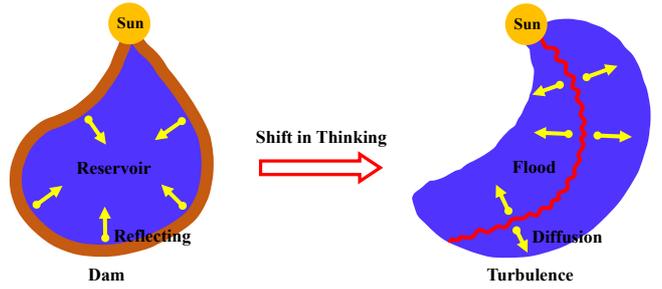}
    \caption{Fundamental shift in thinking and paradigm for
understanding energetic particles and magnetized plasmas in magnetic
turbulence. Conventional paradigm of ``reservoir'' concept on the
left: the hypothetical outer reflecting boundaries or diffusion
barriers (i.e., deterministic structures at macroscale) play the
role of ``reservoir dam'' to inhibit, reflect, and redistribute
energetic particles to form the uniform distribution. Novel paradigm
of ``flood'' concept on the right: the parallel and perpendicular
diffusion processes (i.e., stochastic transport at microscale)
caused by the ubiquitous magnetic turbulence effectively and
uniformly distribute the energetic particles in radial direction and
in longitude and latitude to form the wide-spread flood-like
distribution in the heliosphere. The authentic shift from
``reservoir'' (left) to ``flood'' (right) offers new insights into
energetic particles and space plasmas in the turbulent magnetic
fields.}
    \label{Shift-in-Thinking}
\end{figure}

The notion of particle reservoirs and the relevant concepts (e.g.,
collimated convection) have been employed for several decades in the
data analyses and interpretations within the observational
community. Essentially, these notions stem from the extreme
assumption of scatter-free transport of particles and invoke an
intuitive hypothesis of deterministic plasma structures at
macroscale, e.g., reflecting boundaries or diffusion barriers
\citep[e.g.][]{Roelof1975,Nolte1975,Roelof1992}. This reservoir
paradigm can be illustrated on the left-hand side of Fig.
\ref{Shift-in-Thinking}. As shown, this popular perception invokes
the existence of hypothetical outer reflecting boundaries or
diffusion barriers in the heliosphere, so as to contain the
particles long enough to uniformly distribute them in
heliolongitude, heliolatitude, and radial distance. In this
paradigm, the conceived outer reflecting boundaries or diffusion
barriers play the central role of ``reservoir dam'' to inhibit,
reflect, and redistribute energetic particles in the so-called
reservoirs. Besides, the reservoir concept completely ignores the
effects of stochastic diffusion and scattering processes, especially
for the cross-field diffusion process. This reflecting and
redistribution process must be quite efficient to quickly dissipate
all the particle intensity gradients (longitudinal, latitudinal, and
radial) in the reservoir. To our knowledge, however, so far no
explicit evidence or quantitative mechanism of such reflecting
boundaries or diffusion barriers has been provided within the
community \citep{He2015}.

According to the traditional pragmatic rationale of SEP event
classification (gradual and impulsive events), the energetic
particles in the so-called impulsive events can only form very
narrow longitudinal and latitudinal distributions in the
heliosphere. However, recent multi-spacecraft observations have
revealed that SEPs released from a very small-sized source (e.g.,
solar flare) can also form the wide-range distribution phenomenon
and the reservoir effect, even in the impulsive or $^{3}$He-rich SEP
events \citep[e.g.][]{Dresing2012,Wiedenbeck2013,Richardson2014}.
Therefore, these new observations question the validity of the
conventional classification paradigm of SEP events. Furthermore, the
observations indicate that the plasma structures at macroscale must
not be the dominant reason causing the formation of the reservoir
effect. Actually, the up-to-date multi-spacecraft observations and
numerical simulations of SEP three-dimensional transport have proven
the violation of the notion of particle reservoirs in the
heliosphere. Instead, we introduce a novel concept and paradigm,
called ``flood'', to much better describe the realistic transport
processes of energetic particles. We point out that the notion of
particle ``flood'' is based solidly on the physically meaningful and
mathematically tractable five-dimensional Fokker-Planck focused
transport model and the relevant numerical simulations. The physical
scenario of the particle ``flood'' is illustrated on the right-hand
side of Fig. \ref{Shift-in-Thinking}. As we can see, the energetic
particles, after being released near the Sun, will diffusively
transport in IMF with turbulence and fluctuations. Due to the
effects of parallel diffusion along and perpendicular diffusion
across the large-scale guide magnetic field, the energetic particles
can transport to distant radial locations and meanwhile to distant
longitudes and latitudes in the heliosphere. The mechanism of
perpendicular diffusion is natural and successful to explain and
reproduce the extended decay phases of SEPs (``reservoirs'')
observed in both gradual and impulsive events. Due to the effect of
perpendicular diffusion, the particles can cross the IMF lines and
transport to distant heliospheric locations, and consequently, the
gradually space-filling phenomenon (``reservoir'') can be naturally
formed. In the scenario of perpendicular diffusion, no artificial
outer reflecting boundaries (diffusion barriers/magnetic mirrors)
need to be invoked to play the role of ``reservoir dam'' to
reflect/redistribute the particles to form the so-called
``reservoir'' phenomenon, especially for explaining the wide-spread
particle distribution of $^{3}$He-rich SEP events often detected in
recent multi-spacecraft observations. In this ``flood'' paradigm,
the stochastic diffusion processes caused by ubiquitous magnetic
turbulence enable the energetic particles to cross the IMF lines and
gradually transport to distant heliospheric locations, analogous to
that the surging dynamical river overflows its banks.

As discussed above, the wide-range flooding (space-filling)
transport of SEPs in the heliosphere is an intrinsic property caused
by stochastic diffusion at microscale rather than deterministic
structure at macroscale. To accurately illustrate the realistic
transport processes of energetic particles, we strongly suggest that
the so-called ``reservoir'' should be renamed ``flood''.
Accordingly, the textbooks/books regarding space physics or solar
physics that contain the term ``reservoir'' must be revised and
updated. The traditional ideas of static structures (including
``reservoirs'') and linear instabilities have been a cornerstone for
many theoretical and observational investigations of space plasmas
\citep[e.g.][]{Matthaeus2011}. Therefore, the renaming and
redefinition (based on stochasticity and nonlinearity) of
``reservoirs'' signifies a fundamental shift in thinking and in
paradigms for studying energetic particles and magnetized plasmas in
turbulent magnetic circumstances. In light of this paradigm shift,
the relevant theories and ideas based on determinism and linearity
in space plasma physics need to be rethought and revised. The new
paradigm of interpretation removes determinism from its central
position in understanding the physical nature of the space plasma
behaviors, phenomena, and effects. Instead, this central role should
be given to stochasticism, especially at microscale in plasmas and
in turbulence. More generally, the philosophical debate between
determinism and stochasticism for understanding the essence of
physics has remained a fundamental and long-standing topic in the
literature. The results in this work, obtained in a quantitative
fashion, favor the explanation of stochasticism (randomness), as the
fascinating nature always reveals to us eventually.

\section*{Acknowledgements}
We were supported partly by the B-type Strategic Priority Program of
the Chinese Academy of Sciences under grant XDB41000000, the
National Natural Science Foundation of China under grants 41874207,
41621063, 41474154, and 41204130, and the Chinese Academy of
Sciences under grant KZZD-EW-01-2. H.-Q.H. benefited from the
partial support of the Youth Innovation Promotion Association of the
Chinese Academy of Sciences (No. 2017091). We acknowledge
SOHO/COSTEP and STEREO/HET for providing the energetic particle data
analyzed in this work.








\appendix

\section{Additional Figures of Simulation Results and Comparisons with Multispacecraft Observations}

\begin{figure}
    \centering
    \includegraphics[width=\columnwidth]{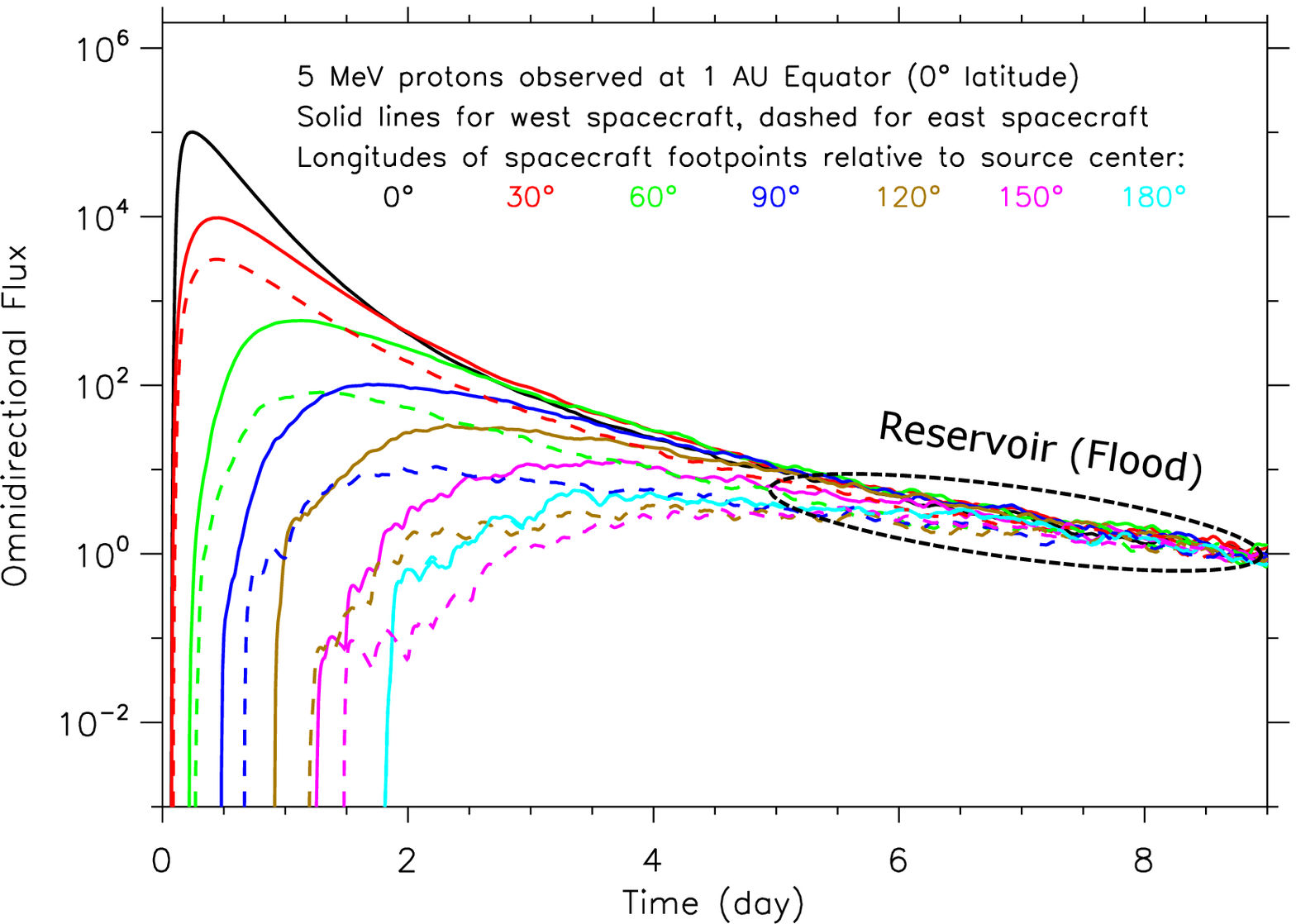}
    \caption{Same as Fig. 1, but for $5$ MeV protons with $\lambda_{r}=0.28$ AU and $\lambda_{x}=\lambda_{y}=0.007$ AU.}
    \label{proton-5MeV-lon}
\end{figure}

The particle reservoir is a universal phenomenon independent of
particle energy. We also investigate the reservoir effect formed by
energetic particles with other energies. Fig. \ref{proton-5MeV-lon}
presents the simulation results of longitudinal transport of $5$ MeV
protons. The diffusion coefficients of $5$ MeV protons are set as
follows: $\lambda_{r}=0.28$ AU (corresponding to
$\lambda_{\parallel}=0.56$ AU at 1 AU), and
$\lambda_{x}=\lambda_{y}=0.007$ AU. Other physical parameters and
conditions are the same as in Fig. 1. Similar to Fig. 1, the
east-west longitudinal asymmetry of particle distribution is also
obviously seen in Fig. \ref{proton-5MeV-lon}. During the late phases
of the energetic particle events, the particle intensities detected
by the widely separated observers display equal intensities and
evolve similarly in time with the same decay rate. Therefore, a
typical particle reservoir spanning $360^{\circ}$ in longitude has
been reproduced in the numerical simulations based on
multidimensional Fokker-Planck transport equation including
diffusion mechanism.

\begin{figure}
    \centering
    \includegraphics[width=\columnwidth]{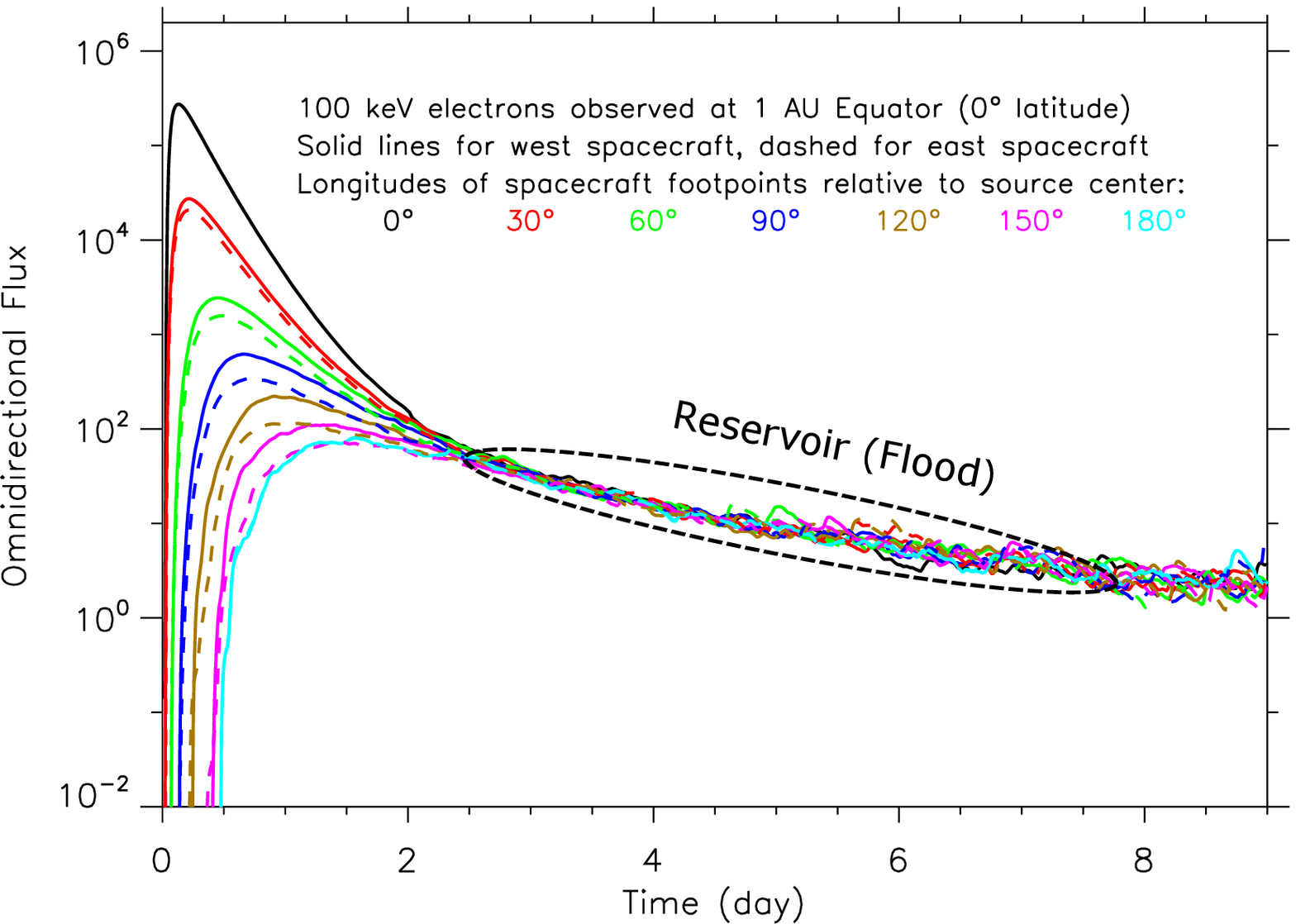}
    \caption{Same as Fig. 1, but for $100$ keV electrons with $\lambda_{r}=0.3$ AU and $\lambda_{x}=\lambda_{y}=0.006$ AU.}
    \label{electron-100keV-lon}
\end{figure}

The heliospheric energetic particle reservoirs are also observed in
electron and heavy-ion data. We also numerically investigate the
reservoirs formed by these particle species. Fig.
\ref{electron-100keV-lon} shows the numerical calculation results of
azimuthal propagation of $100$ keV electrons with diffusion
coefficients as follows: $\lambda_{r}=0.3$ AU (corresponding to
$\lambda_{\parallel}=0.6$ AU at 1 AU), and
$\lambda_{x}=\lambda_{y}=0.006$ AU. Other physical parameters and
conditions are the same as those in Fig. 1. As we can see in Fig.
\ref{electron-100keV-lon}, the distribution of energetic electrons
clearly presents the feature of east-west azimuthal asymmetry.
During the decay phase of the energetic electron events, the
electron fluxes observed by the very widely azimuthally separated
spacecraft present equal fluxes and decrease similarly in time
within the range of statistical errors. Hence, a typical energetic
electron reservoir spreading across $360^{\circ}$ in heliolongitude
has been reproduced in the large-scale simulations based on
Fokker-Planck transport equation incorporating particle diffusion
processes.

\begin{figure}
    \centering
    \includegraphics[width=\columnwidth]{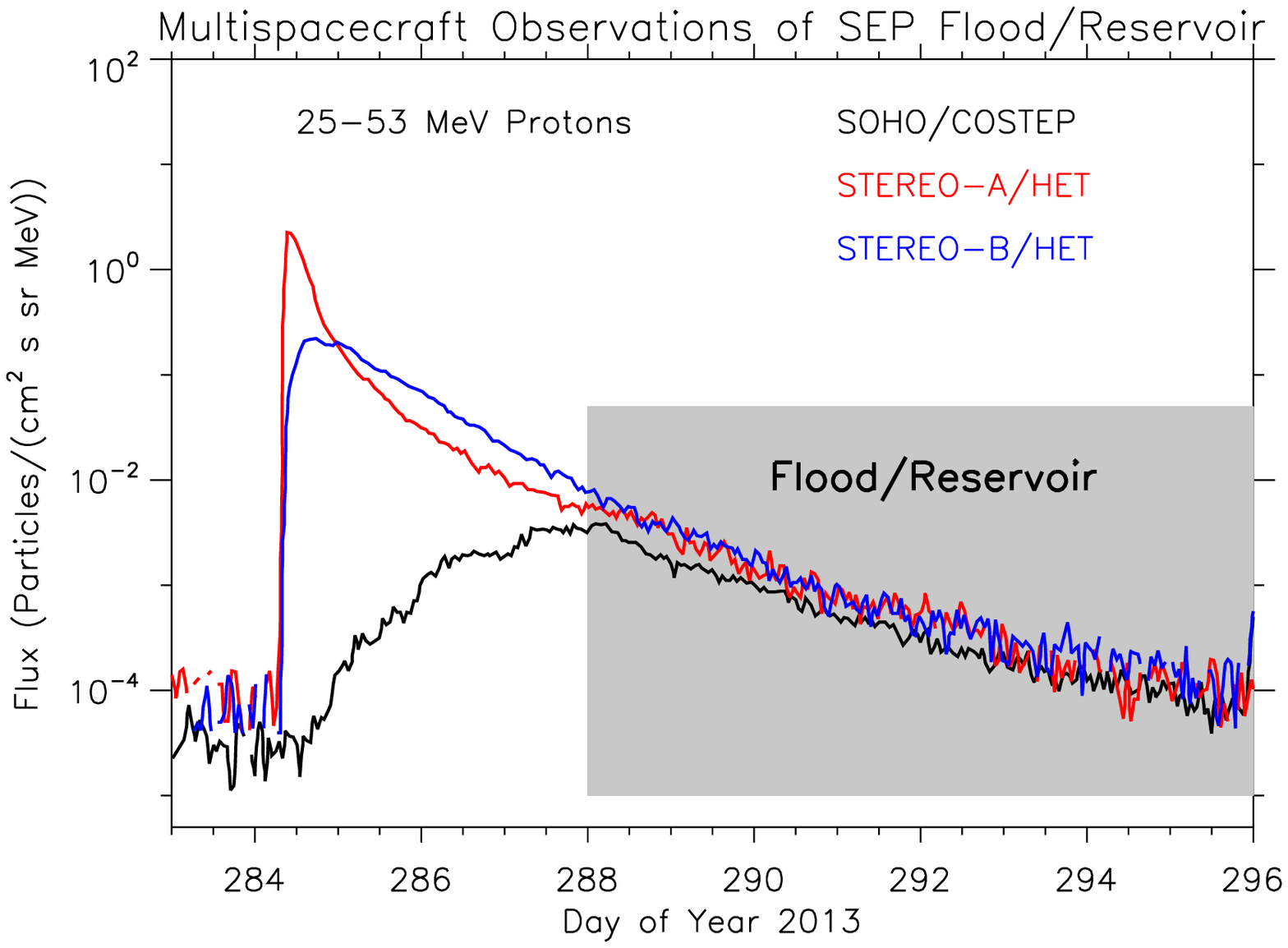}
    \caption{Multispacecraft observations (SOHO, black; STEREO-A, red; STEREO-B, blue) of flood/reservoir effect (denoted by the gray shaded area) in the 2013 October 11 (DOY 284) SEP event.}
    \label{Reservoir-observations}
\end{figure}

The longitudinal flood/reservoir effect has been widely investigated
with multispacecraft measurements (e.g., SOHO, STEREO-A, STEREO-B,
ACE, Wind, etc.) in the observational community. Undoubtedly, the
multispacecraft observations have provided important information on
the transport and distribution of SEPs in the heliosphere. Fig.
\ref{Reservoir-observations} displays the multispacecraft detections
of the flood/reservoir phenomenon observed by SOHO and the twin
spacecraft STEREO-A and STEREO-B during the 2013 October 11 (DOY
284) SEP event \citep[adapted from][]{Anastasiadis2019}. At the
onset of this SEP event, the longitudes of spacecraft SOHO,
STEREO-A, and STEREO-B in the Heliographic Inertial (HGI) Coordinate
System are $302^{\circ}$, $89^{\circ}$, and $162^{\circ}$,
respectively. The solar eruptive activity associated with this SEP
event occurred at E104 in longitude as seen from Earth. In Fig.
\ref{Reservoir-observations}, the black, red, and blue curves denote
the time-flux profiles of 25-53 MeV protons in this SEP event
measured by SOHO, STEREO-A, and STEREO-B, respectively. As we can
see, the time-flux profiles observed at the three widely separated
spacecraft display quite different shapes during the onset and early
phases of the SEP event. However, after 2013 October 15 (DOY 288),
all the SEP fluxes detected at the three locations present
approximately the same intensities and gradually decline to lower
levels with nearly the same decay rate, which is the typical feature
of the SEP flood/reservoir effect. The flood/reservoir phenomenon in
this SEP event lasted for nearly 8 days (denoted by the gray shaded
area). Through qualitative comparisons between observation results
in Fig. \ref{Reservoir-observations} and simulation results in Figs.
\ref{proton-5MeV-lon} and \ref{electron-100keV-lon} and in Fig. 1 in
the main text, we can readily see that all the essential features of
flood/reservoir effect in SEP events have been numerically
reproduced in the three-dimensional transport simulations without
invoking the hypothesis of outer reflecting boundaries or diffusion
barriers. Therefore, the conventional hypothesis of reflecting
boundaries or diffusion barriers needs not to be necessarily invoked
in the numerical reproduction of SEP flood/reservoir effect. In
addition, note that the numerical reproduction results of SEP
floods/reservoirs will remain qualitatively unaffected by the
varying values of the model parameters in the three-dimensional
transport simulations. We can further conclude that the stochastic
diffusion processes (both parallel and perpendicular) at microscale
in magnetic turbulence play a crucial role in the formation of
particle floods/reservoirs and other relevant phenomena in space and
astrophysical plasmas.

\section{Remarks}

The gap of understanding between the observational and theoretical
communities persists in space science for quite a long time. For
simplicity in analyzing and interpreting data, the authors in the
observational community usually employ some specific assumptions,
such as ``collimated convection'' or scatter-free propagation
(deterministic). Undoubtedly, these simplified scenarios can be
tackled more easily. However, the analysis results obtained may
significantly deviate from the realistic physical mechanisms behind
the observations and the phenomena. The solar wind and the
particles, fields, waves, and structures immersed in it naturally
constitute a highly turbulent complex system. In principle,
statistical descriptions based on nonlinearity and stochasticity are
essential for the studies of space plasma physics. Therefore, it is
quite pressing to bridge the wide gap between the theoretical and
observational communities, within which there exist somewhat
different terminologies (some misnomers) for the same phenomena or
effects. In this work, we have demonstrated this important issue
with a typical example, i.e., the so-called particle reservoir
(flood) effect or phenomenon. In the future, we shall pay attention
to and investigate other space plasma behaviors, phenomena, and
effects.

\bsp    
\label{lastpage}
\end{document}